\documentstyle[11pt,ska,twoside,psfig]{article}
\begin{document}
\title{Transient phenomena}
 \author{G.~Woan}
\affil{Department of Physics and Astronomy\\University of
Glasgow\\Glasgow,~G12~8QQ}
\begin{abstract}
The SKA's design is driven by the needs of cutting-edge radio
astronomy for sensitivity and spacial resolution.  However its
design is also driven by the desire to explore the transient radio
Universe. In addition to pulsars, the SKA will be able to carry
out high time resolution observations of several classes of known
and predicted transient sources.  Here we consider a selection of
them, and describe how observational demands affect the
instrument's design.
\end{abstract}
\section{Introduction}
Two of the most famous examples of serendipity in astronomy
resulted from chance observations of transient phenomena. Gamma
ray bursts were first identified in the late 1960s and early 70s
by the Vela satellite system, launched to verify Soviet adherence
to the nuclear test ban treaty (Klebesadel et al.\ 1973), and at
about the same time a Cambridge radio telescope designed for
interplanetary scintillation studies resulted in the discovery of
pulsars (Hewish et al.\ 1968).  By definition, serendipity is not
something one can rely on,  but both these examples were a result
of observations that explored a region of the spectrum with
unprecedented sensitivity and time resolution.  These are
fundamental features of the SKA design, and we can certainly hope
it will deliver something equally unexpected. However, we can see
already that our understanding of many well-known transient
sources will be enormously improved by SKA data.  In addition to
good sensitivity and time resolution, the SKA will offer the
ability to carry out agile, event driven, observations and regular
all-sky surveys -- a feature normally associated only with
low-frequency instruments.

It will be possible to map the structural evolution of transient
sources with the SKA (a MERLIN example is considered in
Section~2.3) but the simple rule we apply to signals of duration
$\tau$ is that they cannot originate from a source larger than
$\sim c\tau$. Although one can imagine pathological coherent
emission propagation processes that would beat this rule, it it
fair to say that the mapping of \emph{very} short timescale
sources will not be possible with the SKA. However astrometry will
be possible, and has a particular significance for transient
sources. For some classes of source we may have just a few seconds
to determine the arrival direction, and suitable observing
programmes for such sources will be discussed later in this
review.

There are two important limiting processes that we should also
keep in mind when considering transient sources.  First,
interstellar scattering becomes increasingly important at low
frequencies and the temporal broadening of transients, effectively
caused by multipath propagation through the turbulent interstellar
medium, scales as $\sim\lambda^4$ (e.g., Cordes 1990).  Very
roughly we would expect a distant galactic or extragalactic
transient to be broadened by $\sim200\,\mu$s at 1\,GHz and
therefore 2\,s at 100\,MHz, although there is significant
variation along different lines-of-sight. Second, because very
short duration sources are necessarily small, they are likely to
be self-absorbed at low frequencies. Both these effects favour
higher frequencies for galactic and extragalactic transient
observations.

\section{Known transient radio sources}
Even excluding pulsars (see Kramer in these proceedings) there is
a remarkably broad range of known or predicted transient sources
and phenomena that have received attention as suitable targets for
SKA science. A brief, and incomplete, list of these includes
\begin{itemize}
\item solar system and extrasolar planetary emission (mostly at
low frequencies, see Farrell et al.\ 1999),
\item solar bursts (Bastian et al.\ 1998),
\item interstellar and interplanetary radio propagation and space
       weather (Lazio 1999; Hick et al.\ 1996),
\item microlensing events (Koopmans and de Bruyn, 1999 and 2000),
\item air-shower events -- radio pulses from cosmic rays hitting the
Earth's atmosphere (Horneffer et al.\ 2002),
\item compact object coalescences (`LIGO events' --  Usov and Katz 2000;
      Hansen and Lyutikov 2001),
\item X-ray binary transients (Fender 1999),
\item gamma ray burst transients and afterglows (Galama and de Bruyn,
      1999; Dado et al.\ 2003),
\item SETI (Welch and Dreher 2000).
\end{itemize}
A review of all these is not possible here, and details of each
are available from the references cited and more general SKA
documentation such as the SKA Working Group 2 Report on transient
phenomena (Lazio et al.\ 2002). However we will consider three
examples in more detail below to highlight the broad range of
phenomena that come under the heading `transients'.
\subsection{Solar imaging}
The SKA will deliver the sub-arcsec resolution, millisecond
sampling and frequency agility necessary to carry out imaging and
3-D tomographic modelling of the solar corona.  These are
techniques currently under development for the proposed
Frequency-Agile Solar Radiotelescope or FASR (Bastian et al.\
1998).  FASR is a $\sim 100$-antenna multi-frequency imaging array
($\sim 0.1$ to 30\,GHz) with high spectral ($\Delta f/f\sim 0.02$)
and temporal ($<1$\,s) resolution.  The SKA will be able to go
even further, investigating the energy release mechanisms in the
corona and the formation and destablisation of large-scale
structures in active regions and coronal loops. Current
techniques, using instruments such as the Nobeyama radioheliograph
at 17\,GHz and the VLA at 5\,GHz have begun to reveal the dynamics
of plasma heating and electron acceleration within the corona on
timescales as short as 0.1\,s (e.g., Gopalswamy 1996; Zhang et
al.\ 2001).
\subsection{Microlensing}
At first sight radio microlensing seems improbable, as most
extragalactic radio sources are too extended to be strongly
microlensed. However the compact features in the jet components of
radiogalxies are sufficiently compact ($\mu$arcsec) to show
microlensing and bright ($\mu$Jy to mJy) to be detected by the SKA
in a matter of minutes (Koopmans \& de Bruyn 1999). These
microlensing transients will occur on timescales of days to weeks
and should be distinguishable from refractive interstellar
scintillation by their different spectral signatures.  A
particularly attractive feature of this technique is that, if the
compact feature is superluminal, the microlensing rate could be as
much as a thousand times greater than that from a stationary
optical source. Koopmans \& de Bruyn (2000) argue that the radio
gravitational lens CLASS B1600+434 shows radio variability
consistent with microlensing by massive compact objects in the
bulge/disk and halo of the lens galaxy at $z=0.4$.  The SKA will
offer the opportunity for more systematic studies of radio
microlensing to  measure the mass function of, and mass
distribution in, high redshift galaxies and investigate the
distribution of dark matter in galaxies as a function of redshift.

\subsection{X-ray binary transients}
X-ray binary systems, comprising an accreting neutron star or
black hole with a donor stellar companion, represent one of the
more accessible classes of exotic sources available to us, and
have received much attention in recent years. As well as radiating
in X-rays and gamma rays these sources can show highly variable
radio emission together with radio jets.  In addition low-mass
X-ray binaries are thought to be relatively strong emitters of
gravitational waves.
\begin{figure}[!]
\centerline{\psfig{figure=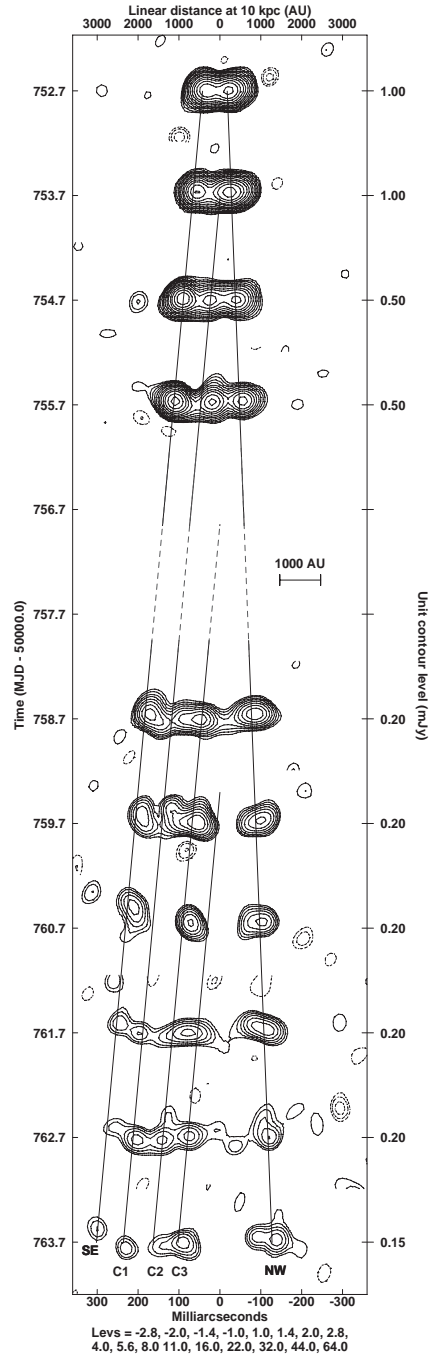,height=18 cm}}
\caption{Radio evolution of the X-ray transient source GRS
1915+104 at 5\,GHz (from Fender et al.\ 1999).  These are MERLIN
maps of total intensity.}
\end{figure}
The transient radio emission from these objects is often dramatic,
and has been followed at many wavelengths (e.g., Brocksopp et al.\
2002).  Typically they comprise a rapid rise in flux, on
timescales of hours and resolved images show rapidly evolving
superluminal radio jets.  A particularly spectacular example was
seen in GRS 1915+105 by Fender et al.\ (1999) using MERLIN
(Fig.~1).

The sensitivity and mapping capability of the SKA will allow the
production of high resolution snapshot images of the evolving
radio jets in these sources. With sensitivities in the $\mu$Jy
range observations can be extended to the extragalactic population
of X-ray binaries in the Local Group (Fender 1999). The mechanism
for the radio jets and the relationship between the X-ray and
radio emission remain unclear, and the SKA will revolutionise the
study of this class of transient source.

\section{Observing programmes and requirements}
The SKA transient programme has a broad remit, and the
observational demands it makes have significant consequences for
the design of the telescope.  Sufficient time resolution is
obviously essential and is common to both the pulsar and transient
programmes. The most stringent constraints for this are from solar
imaging, for which the SKA should be able to deliver snapshot
images of coronal features on timescale of milliseconds.

Transient observations will sometimes be triggered by observations
from other instruments. For GRBs we have some dispersive delay
between the gamma ray and radio signals, relaxing the response
time requirements slightly, but there will be other situations
when transient observations are totally missed in real-time. To
recover the data would need a `delay buffer' in the SKA design,
capable of retrospective beam forming and synthesis.  Implementing
such a buffer might be difficult, but would greatly enhance the
SKA's transient performance.  In addition to follow-up and
triggered response observations the SKA will be configured to
perform directed searches for transients, such as from X-ray
binaries or from extreme scattering events (Fiedler et al.\ 1987).
In addition, an efficient SETI monitoring programme can be
developed by synthesising multiple beams directed at local stars.
The multiple beam-forming capabilities of the SKA can also be
exploited to carry out the first systematic all-sky transient
surveys, using sub-arrays to survey different parts of the sky and
distinguish between RFI and true transients (Lazio at al.\ 2002).

These transient programmes will span the full frequency range of
the strawman design (150\, MHz to 15\,GHz).  The low frequency end
will be particularly important in planetary observations, galactic
and extragalactic giant pulse observations in the pulsar transient
programme and possibly for follow-up observations from future
gravitational wave inspiral detections. Higher frequency
observations are less affected by self-absorbtion and angular and
temporal broadening by the ISM, particularly towards the galactic
centre. The field-of-view (FoV) requirements are also stringent if
the whole sky is to be covered in a reasonable length of time.  A
full-sky survey taking 1 day, with a 1 square degree FoV leaves
only  5\,s of integration per pointing. The astrometry
requirements for transients are also demanding, with 1 to 10\,mas
resolution at 5\,GHz desirable if the SKA is to compliment ALMA
and the NGST.

The SKA promises to be a remarkable instrument, and all those
involved with its conception are keen to explain how it will
revolutionise our existing understanding of the radio Universe.
However its most lasting contribution may well be in the
unexpected, previously unimagined, areas of astrophysics that it
will reveal. The transient programme in particular covers volumes
of parameter space in sensitivity and time resolution that are far
in excess of anything that has been available before, and we await
its exploration with much excitement.
%

\end{document}